\documentclass[aps,prl,twocolumn,superscriptaddress,nobibnotes,longbibliography,10pt]{revtex4-2}
\usepackage{graphicx}
\usepackage{dcolumn}
\usepackage{bm}
\usepackage{makecell}
\usepackage[utf8]{inputenc}
\usepackage[T1]{fontenc}
\usepackage{mathptmx}
\usepackage{etoolbox}
\usepackage{xcolor}
\usepackage{float}
\usepackage{amsmath}
\usepackage{amssymb}
\usepackage{hyperref}
\usepackage{comment}
\usepackage[normalem]{ulem} 

\begin{document}

\title{Secure Quantum Key Distribution Using a Room-Temperature Quantum Emitter} 

\author{Ömer S. Tapşın}
\thanks{These authors contributed equally}
\affiliation{QLocked Technology Development Inc., İzmir, 35430, Turkey}

\author{Furkan Ağlarcı}
\thanks{These authors contributed equally}
\affiliation{QLocked Technology Development Inc., İzmir, 35430, Turkey}
\affiliation{Department of Physics, İzmir Institute of Technology, İzmir, 35430, Turkey}

\author{Roberto G. Pousa}
\affiliation{ICFO - Institut de Ciencies Fotoniques, The Barcelona Institute of Science and Technology, 08860 Castelldefels, Barcelona, Spain}
\affiliation{SUPA Department of Physics, University of Strathclyde, John Anderson Building, 107 Rottenrow East, Glasgow, G4 0NG, UK}

\author{Daniel K. L. Oi}
\affiliation{SUPA Department of Physics, University of Strathclyde, John Anderson Building, 107 Rottenrow East, Glasgow, G4 0NG, UK}

\author{Mustafa Gündoğan}
\affiliation{Institut f\"{u}r Physik, Humboldt-Universit\"{a}t zu Berlin, Berlin, 12489, Germany}

\author{Serkan Ateş}
\email[]{Corresponding author: serkanates@iyte.edu.tr}

\affiliation{QLocked Technology Development Inc., İzmir, 35430, Turkey}
\affiliation{Department of Physics, İzmir Institute of Technology, İzmir, 35430, Turkey}

\begin{abstract}
\textcolor{black}{On-demand generation of single photons from solid-state quantum emitters is essential to build practical quantum networks and QKD systems by \textcolor{black}{potentially} enabling higher secure key rates (SKR) and lower quantum bit error rates (QBER) \textcolor{black}{in short-range distances}. Room-temperature operation is particularly important as it eliminates the need for bulky cryogenic setups, reducing complexity and cost for real-world applications. In this work, we showcase the versatility of defects in hexagonal boron nitride (hBN) at room temperature by implementing the B92 protocol.} Our experiments yield a sifted key rate (SiKR) of 17.5~kbps with a QBER of 6.49\% at a dynamic polarization encoding rate of 40~MHz, and finite-key analysis provides a SKR of 7 kbps, one of the highest achieved for a room-temperature single photon source. \textcolor{black}{We analyzed the non-decoy efficient BB84 using our hBN emitter and other promising quantum dot source for QKD, and compare their key performance with a single quantum repeater scenario.} We also explore potential applications of hBN defects beyond QKD and analyze scenarios that could outperform conventional point-to-point QKD schemes. These results underscore the promise of hBN emitters for advancing quantum communication technologies.
\end{abstract}


\maketitle 

\section*{Introduction}

Among various quantum technologies, quantum key distribution (QKD) stands out as one of the most mature and well-established applications, laying the foundation for future quantum networks and distributed quantum computing~\cite{Wehner.Hanson.2018,Lu.Pan.2021,Gyongyosi.Imre.2022}. By leveraging fundamental quantum principles such as the no-cloning theorem, nonlocality, and the uncertainty principle, QKD enables secure shared random key generation between remote parties. QKD protocols are typically categorized into prepare-and-measure (PM) schemes~\cite{Bennett.Brassard.2014,Bennett.Bennett.1992,Bruß.Bruß.1998,Inoue.Yamamoto.2002} and entanglement-based schemes~\cite{Ekert.Ekert.1991,Bennett.Mermin.1992}. In PM QKD, weak coherent pulses (WCPs) are commonly used to generate flying qubits as quantum information carriers. \textcolor{black}{However, their multi-photon emissions are exploitable by an eavesdropper through photon number splitting (PNS) attacks, which compromise the security of the protocol in lossy channels. This vulnerability, coupled with the limited source brightness, further constrains the achievable key generation rate}. Decoy-state methods address these challenges by mitigating the risks of multi-photon emissions, albeit at the cost of increased protocol complexity~\cite{Hwang.Hwang.2003,Lo.Chen.2005,Wang.Wang.2005}. Alternatively, deterministic single-photon generation by quantum emitters provides an on-demand solution for practical quantum communication, allowing extended transmission distances with enhanced security~\cite{Xu.Pan.2020,Couteau.Weihs.2023iul}. 

Semiconductor quantum dots (QDs), as one of the most studied quantum emitters, have demonstrated significant potential for PM QKD due to their high purity and brightness~\cite{Waks.Yamamoto.2002vsu,Zahidy.Midolo.2024,Zhang.Pan.2024}, as well as for entanglement-based QKD through cascaded exciton-biexciton emission~\cite{Dzurnak.Shields.2015,Basset.Trotta.2021,Schimpf.Rastelli.20214yr,Basset.Trotta.2023}. Tremendous efforts have been made to enhance these key properties of QDs~\cite{Lodahl.Stobbe.2013,Arakawa.Holmes.2020}. However, the implementation of QD sources in QKD requires significant growth infrastructure and cryogenic environments, which introduces considerable complexities~\cite{Vajner.Heindel.2022}. In contrast, quantum emitters that operate efficiently at room temperature have garnered considerable attention. For example, fiber-based QKD has been demonstrated using single photons generated in the telecom O-band from defects in GaN~\cite{Zhang.Ling.2024}. Similarly, defects in diamond emitting in the visible band have been employed for free-space QKD~\cite{Beveratos.Grangier.2002, Alléaume.Grangier.2004, Leifgen.Benson.2014}. In addition to these materials, defects in hexagonal boron nitride (hBN) are attracting increasing attention due to their efficient single-photon generation at room temperature~\cite{Tran.Aharonovich.2016}, their 2D nature, and their ease of integration with quantum photonic and plasmonic structures~\cite{Kianinia.Aharonovich.2022,Montblanch.Ferrari.2023}. Thanks to its wide bandgap, hBN hosts several optically active defects that emit over a broad spectral range, from the visible (VIS) to near-infrared (NIR)\cite{Cholsuk.Vogl.2024}, enabling a diverse range of applications in quantum technologies\cite{Cakan.Vogl.2024}. The potential of hBN defects has been demonstrated as single-photon sources in QKD implementations, successfully employing the B92~\cite{Samaner.Ateş.2022} and BB84~\cite{Al‐Juboori.Aharonovich.2023} protocols in free-space, and leading to their inclusion in an upcoming space-based mission~\cite{Quick32024}. \textcolor{black}{Furthermore, the recently discovered electronic spin~\cite{Stern2022.Atature2024,Durand.Jacques.2023} associated with the optical transition opens up the potential for developing spin-photon interfaces that could serve as quantum registers for quantum networking.}

\begin{figure*}[!htp]
    \centering
    \includegraphics[width=0.95\textwidth]{Fig_1.pdf}
    \caption{Schematic diagram of the experimental configuration employed for the optical investigation of defects in hBN and the implementation of B92 QKD protocol. The Alice module on the left integrates the RT-SPS box, which includes a $\mu$-PL setup used for excitation of defects and collection of emission for spectral analysis. It also includes HBT box, which illustrates the Hanbury-Brown and Twiss interferometer to demonstrate the single photon nature of the collected emission. Alice implements a resonant electro-optic modulator (EOM) for the polarization encoding of individual photons, while the Bob module on the right randomly measures the photons having non-orthogonal polarization states prepared by Alice.}
    \label{fig:1}
\end{figure*}

All QKD demonstrations mentioned above rely on the active polarization encoding of single photons using electro-optic modulators (EOMs). Practical QKD would benefit from a bright, room-temperature single-photon source and a high-speed polarization modulator. Fiber-based QKD with defects in GaN leverages commercially available high-speed electro-optic phase modulators operating at telecom wavelengths~\cite{Zhang.Ling.2024}. In contrast, free-space QKD demonstrations with defects in diamond and hBN are constrained by free-space EOMs operating in the visible spectrum. These modulators are limited to practical speeds of 1 MHz due to their high voltage requirements for polarization modulation~~\cite{Leifgen.Benson.2014, Samaner.Ateş.2022,Al‐Juboori.Aharonovich.2023}, although the emitters can support much higher speeds due to their short lifetime. 

Here, we present, to the best of our knowledge, the fastest dynamically modulated QKD system based on the B92 protocol~\cite{Bennett.Bennett.1992}, utilizing polarization-encoded single photons emitted from a single defect in hBN at a 40~MHz clock rate. Additionally, temporal filtering based on the emitter's decay time is employed to optimize the quantum bit error rate (QBER) and the sifted key rate (SiKR)~\cite{Kupko.Heindel.2020}, alongside performing both asymptotic~\cite{Waks.Yamamoto.2002,Gottesman.Preskill.2004} and finite-key analysis~\cite{Mafu.Petruccione.2013,Pousa.Jeffers.2024}. \textcolor{black}{We will conclude the paper with an analysis of the performance requirements for a potential spin-photon interface based on hBN to function as a quantum repeater node capable of surpassing the direct transmission limit.}

\section*{Optical Characterization of the Single Photon Source}
Figure~\ref{fig:1} shows the detailed experimental setup used for the optical excitation of defects, the analysis of the emission spectrum, the verification of the single-photon nature of the emission, and the execution of the QKD protocol. The sample is excited with a 483~nm pulsed laser using a high numerical aperture objective (NA~=~0.9), which is also used for the efficient collection of emission. Additionally, a Hanbury-Brown and Twiss interferometer equipped with two single-photon detectors with low dark counts (40~Hz) is employed to perform photon correlation experiments on the spectrally filtered emission from the defect. For the QKD demonstration, the filtered single-photon emission is guided to the polarization encoding part via a polarization-maintaining fiber (PMF). The details of the QKD scheme and setup are described below.

Multilayer hBN flakes (obtained from Graphene Supermarket) dropcasted on a $SiO_2/Si$ substrate are used as the material system. Photoluminescence (PL) spectrum of the defect investigated in this work is shown in \autoref{fig:2}(a). The sharp peak at 626~nm is identified as the zero-phonon line (ZPL), while the weaker and broader signals around 680~nm correspond to the phonon sidebands (PSB). The energy difference between ZPL and PSB matches very well with the energy of the high-energy Raman-active and infrared optical phonon modes of hBN~\cite{Cuscó.Artús.2016}. The inset of \autoref{fig:2}(a) shows the linear polarization behavior of the ZPL emission with 70\% visibility~\cite{Kumar.Vogl.2024}. Time-resolved PL spectroscopy is employed on the ZPL emission to determine its decay time. Figure~\ref{fig:2}(b) shows the result of the experiment with a lifetime of 4.58~ns obtained from a single-exponential decay model. Finally, to demonstrate the single-photon nature of the collected emission, a photon correlation experiment is performed on the spectrally filtered ZPL emission under a 40~MHz excitation rate at saturation, which also represents the experimental conditions for the QKD demonstration, as discussed later. The inset of \autoref{fig:2}(b) shows the histogram of correlations with an antibunching value of $g^{(2)}(0)~=~0.24$ at zero delay, normalized to the correlations at long time delays. The deviation from the ideal value of $g^{(2)}(0)~=~0$ is mainly attributed to the imperfect spectral filtering of the emission using a bandpass filter with a 10~nm FWHM around the ZPL wavelength. The purity of the emission can further be enhanced via post-growth processes, such as thermal annealing~\cite{Li.Aharonovich.2019} that also narrows the spectral width. The bunching behaviour around zero time delay is due to the three-level structure of the defect~\cite{Martinez.Jacques.2016}.

\begin{figure}[!tp]
    \centering
    \includegraphics[width=0.45\textwidth]{Fig_2.pdf}
        \caption{Optical characterization of investigated defect (a) PL spectrum of the defect showing a ZPL emission at 626 nm and phonon sidebands near 680 nm. Inset: linear polarization of ZPL emission with 70\% visibility. (b) Time-resolved ZPL emission from the defect, where a single-exponential decay model yields a decay time of 4.58~ns. Inset: second-order correlation measurement under pulsed excitation at 40~MHz, with an antibunching value of $g^{(2)}(0)~=~0.24$, normalized to the correlations at long delays.}
    \label{fig:2}
\end{figure}

\section{Free-Space Quantum Key Distribution}
\textcolor{black}{Despite its drawbacks compared to BB84---namely lower loss and error tolerances and greater vulnerability to certain attacks---we use the B92 protocol for practical, experimental reasons, primarily to showcase the versatility of hBN emitters at room temperature. In the B92 (or two-state) protocol, QKD is implemented between Alice (transmitter) and Bob (receiver).} As shown in \autoref{fig:1}, spectrally filtered single photons are sent via a PMF, and Alice actively encodes the polarization states \({|V\rangle, |D\rangle}\), corresponding to the linear and diagonal bases \({\mathbb{X}, \mathbb{Z}}\) which map to bits \({0, 1}\).

The protocol is implemented in the following manner. Alice generates 20~MHz and 40~MHz signals that are synchronized with the FPGA clock of the arbitrary waveform generator. The 40~MHz signal is used to trigger the single-photon generation, while the 20~MHz control signal is used to drive the resonant EOM for dynamic polarization manipulation of the single-photons. The EOM is driven by 7 $V_{pp}$, such that polarization encoding is achieved by overlapping the control (20~MHz) and trigger (40~MHz) signals in a structured pattern, alternating between high and low voltage states, represented as a "1-0-1-0..." sequence. In addition, passive-polarization components are employed by Alice for state preparation. Before the EOM, a linear polarizer (P) and a quarter-wave plate (QWP) are used to first select the vertical polarization state and convert it to circular polarization, which is then modulated by the EOM, resulting in elliptically polarized light. A second QWP is used after the EOM to linearize the polarization state to -22.5° and 22.5°. A half-wave plate (HWP) is then used to rotate the reference frame to acquire non-orthogonal states, corresponding to vertical and diagonal (45°) polarization, providing $\pm$3.5~V. Subsequently, the prepared states are transmitted to Bob over the free-space quantum channel, where a 50:50 beam splitter (BS) is first used for the random selection of the measurement basis. The polarizing beam splitter (PBS) in the transmission path represents the linear basis (${\mathbb{X}}$), while the reflection path, along with the HWP and PBS, represents the diagonal (${\mathbb{Z}})$ basis, corresponding to bits ${1, 0}$, respectively. Photon detection is carried out by fiber-coupled single-photon detectors, which are also used for the measurement of second-order photon correlation function, as described above.

The QKD experiment was performed with spectrally filtered ZPL emission. The brightness of the emitter was characterized based on a measured total count rate of 851~kHz using a single APD at the output port of the HBT setup. Taking into account all losses in the $\mu$PL setup and a 70\% detector efficiency (Excelitas AQRH-16), a count rate of 2.085~MHz was obtained after the microscope objective, indicating an overall collection efficiency of 5.2\%. A critical parameter for QKD is the mean photon number, $\mu$, before the quantum channel. Considering the efficiency of the $\mu$PL setup, the PM fiber coupling, and the losses in the polarization encoding part of Alice's system, the efficiency of the transmitter, $\eta_{tran}$, was calculated to be 0.252, yielding a mean photon number, $\mu$, of 0.0131. On Bob's side, the system is relatively simple, consisting of basic optics and two single-photon detectors, which resulted in a receiver efficiency, $\eta_{rec}$, of 0.42. This includes detector efficiency, transmission of the optics, and coupling into the multimode fibers used before APDs. The efficiency of the QKD setup, from the microscope objective to the single-photon detectors in the receiver, was estimated by measuring each component individually with an attenuated laser at the emission wavelength.

\begin{figure*}[!htp]
    \centering
    \includegraphics[width=0.95\textwidth]{Fig_3.pdf}
    \caption{Results of QKD experiment \textcolor{black}{with the B92 protocol}. (a) Normalized histogram of APD detection events on Bob module. The red solid line on the histogram is the decay curve of the ZPL emission. The unshaded region shows the start ($t_0$) and the width ($\Delta$t) of temporal filtering. (b) A two-dimensional representation of the sifted key rate, (c) quantum bit error rate, and (d) secure key rate as a function temporal filtering parameters $t_0$ and $\Delta$t extracted from the measured QKD data. Vertical dashed lines on each map indicate the optimal temporal filtering parameters for maximizing SKR. (e) SiKR, QBER, and SKR results for the indicated dashed lines. \textcolor{black}{We follow the method presented in~\cite{Mafu.Petruccione.2013} for these calculations as explained in the text.}}
    \label{fig:3}
\end{figure*}

To obtain the QKD parameters, the sifted key rate and the quantum bit error rate (QBER), the trigger signal at 40~MHz is recorded alongside the APD detections at Bob. This enables sifting and QBER calculations to be performed over the classical channel. Time tags corresponding to QKD events are selected based on the recorded trigger signal, and double-detection events as well as empty pulses are discarded during sifted key generation at Bob. Additionally, a periodic bit sequence of "1-0-1-0..." is generated in correspondence with the trigger signal, representing the original bit sequence sent from Alice to Bob. The clicks from APD-1 and APD-2, which are used for the bit sequences of 1's and 0's measured by Bob, are then compared to Alice's bit sequence to extract the QBER. 
 
Figure~\ref{fig:3}(a) shows the normalized histogram of APD detection events at Bob, together with the decay curve of the emission obtained from the time-resolved PL measurement. As observed, while the single-photon emission is stronger in the early part of the detection window, the weaker signal at later times leads to a reduced signal-to-noise ratio. As discussed earlier, linearly polarized single-photon emission is used as the source for the QKD process. However, the degree of linear polarization strongly depends on the decay time of individual single photon generation~\cite{Kumar.Vogl.2024}, which consequently affects the generated bit rate and the QBER. Additionally, the dark counts distributed across the detection window play a significant role in determining the QBER. To optimize the key rate and QBER in relation to the emitter's decay time, a temporal filtering process is applied~\cite{Kupko.Heindel.2020}. The unshaded region shown in \autoref{fig:3}(a) represents the window used for temporal filtering, characterized by a start time $t_0$ with respect to the synchronized trigger signal and a width $\Delta t$. By varying $\Delta t$ between 3 and 12~ns for each $t_0$ from 0 to 4~ns (in 100~ps steps), the sifted key rate (SiKR) and QBER are calculated from the measured raw key, as shown in \autoref{fig:3}(b). The high-speed operation of the QKD system enables the generation of a SiKR up to 17.5~kbps. Noting that the efficiency of the B92 protocol is half that of BB84, the observed sifted key rate (SiKR) is, to the best of our knowledge, the highest achieved from a room temperature single-photon source (SPS) with active polarization encoding\cite{Beveratos.Grangier.2002,Leifgen.Benson.2014,Samaner.Ateş.2022,Al‐Juboori.Aharonovich.2023,Zhang.Ling.2024}. In addition to SiKR, the effect of temporal filtering on QBER is demonstrated in \autoref{fig:3}(c). 

Finite key analysis for the B92-protocol~\cite{Mafu.Petruccione.2013} is performed for secure key rate (SKR) calculations, as described in \autoref{met:finite}, considering the same temporal filtering parameters used for SiKR and QBER calculations from raw data. For this purpose, leak estimation is carried out following the one-way error reconciliation method~\cite{Tomamichel.Elkouss.2017} such that the secure key length $l$ is bounded by,
\begin{equation}
    \resizebox{0.43\textwidth}{!}{$
    l \leq \displaystyle N_{R}[1-H_{min}(X_A|E)] - L_{EC} - 2 \log_2 \left( \frac{1}{2 \epsilon_{PA}} \right) - \log_2 \left( \frac{2}{\epsilon_{cor}} \right)
    $}
    \label{eq:main}
\end{equation}

Here, $N_R$ is the number of total measured events per second, $H_{min}$ is the minimum entropy, $L_{EC}$ is the leak bits during error reconciliation, $\epsilon_{PA}$ is the privacy amplification failure probability and $\epsilon_{cor}$ is the correctness failiure probability. Figure~\ref{fig:3}(d) shows the 2D map of the calculated SKR for the temporal filtering parameters given above using \autoref{eq:main}. The dashed lines on the plots highlight the optimal temporal filtering parameters, determined to be ($t_0$,~$\Delta t$)~=~(0.5~ns, 3–10.5~ns) for achieving the best SiKR, QBER and SKR values. As shown in \autoref{fig:3}(e), a QBER of 6.49\% is achieved for the highest SiKR of 17.5~kbps and SKR of 7~kbps under the given temporal filtering conditions. This represents one of the highest reported SKR obtained from a QKD system with active modulation of polarization states from a room-temperature single-photon source. The abrupt changes in the behavior of SiKR, QBER, and SKR at 10.5~ns, observed in \autoref{fig:3}(e) (shaded region), are attributed to the pulse shape of the 40~MHz trigger signal, which has a roughly 10~ns flat region after the rising and falling edges. In this context, temporal filtering helps select the optimal portion of the data for the calculation of QKD parameters within the flat region of the trigger signal, in addition to the conventional purpose of signal-to-noise improvement, as previously reported~\cite{Kupko.Heindel.2020, Samaner.Ateş.2022}. Therefore, to fully exploit the performance of a single-photon source with a slow decay time, the shape of the pulse driving the EOM must be optimized such that the flat region of the pulse is wider than the decay time. Achieving this condition is not trivial under high-speed modulation conditions. 

\begin{figure}[!tp]
    \centering
\includegraphics[width=0.485\textwidth]{Fig4_updated.pdf}
    \caption{
    \textcolor{black}{Secure key rates relative to channel loss (fibre distance calculated using 0.2 dB per km) are presented here for the non-decoy BB84 protocol, optimising both basis bias ($p_x$) and Alice's source pre-attenuation for each dB loss. These keys are estimated for three quantum emitters, the green lines illustrate the expected performance of our emitter, the blue lines show the improved performance assuming cavity coupling and minimized experimental losses, highlighting the potential of the isolated defect studied in this work, and the black lines represent an experimental QD with promising characteristics for QKD ~\cite{Zhang.Pan.2024}, i.e. low-$g^{(2)}(0)$ and high mean photon number (black lines). Furthermore, we also show attainable SKRs with a single-node quantum repeater scheme that utilizes two quantum memories with memory times ($T_2$) of 5~ms (dotted red line) and 10~ms (dash-dot red line) with optimized node position in the asymptotic regime~\cite{Luong.Lutkenhaus.2016}. In the finite-key regime, we define the block size as the number of sent signals by Alice $N_{S, BB84} = R \, t_S$, where $R$ is the excitation rate and $t_S$ is the acquisition time. We analyse finite blocks of 1 (dotted line), 10 (dashed line) or 100 seconds (dot-dash line) and the asymptotic limit (solid lines). However, these times does not generate the same block size for each QD since they operate at different excitation rates, see Table \ref{table:1}. }}
    \label{fig:4}
\end{figure}

\textcolor{black}{Here, we should note that} \textcolor{black}{the unconditional security of the B92 protocol relies on perfect positive-operator-valued measurements (POVMs), but practical systems}, \textcolor{black}{such as this experiment,} \textcolor{black}{only use two projective measurements. This limitation weakens security, allowing Eve to exploit imperfections through unambiguous state discrimination (USD) measurements \cite{Duek2000}. Although some countermeasure strategies against USD attacks for the B92 protocol exist in the literature \cite{Ko2017}, implementing them within a lossy channel framework with realistic parameters remains challenging.} \textcolor{black}{On the other hand, unlike B92, the BB84 protocol, with its larger state set, is more resilient to such attacks and offers well-established frameworks for lossy channels \cite{Tamaki2004}.}  \textcolor{black}{Therefore, to assess the platform's capabilities beyond B92, we evaluate its performance under hypothetical BB84 conditions, which offer improved security and advantages as explained above. In order to do so}, we benchmark our RT source using asymptotic~\cite{Waks.Yamamoto.2002,Gottesman.Preskill.2004} and finite key frameworks~\cite{Morrison.Fedrizzi.2023,Pousa.Jeffers.2024} for the BB84 protocol, with the experimental parameters given in \autoref{table:1}(see \autoref{met:met}), against channel loss. The solid green line in \autoref{fig:4} shows the result of the asymptotic key rate analysis with a maximum tolerable loss of \textcolor{black}{24~dB}. The dashed green lines represent the secure key rate obtained with the finite key analysis based on multiplicative Chernoff bounds, by estimating lower bounds of the \textcolor{black}{sent} non-multiphoton events ($\underline{N}_{S,nmp}^{X,Z}$) and upper bound of the phase error rate ($\bar{\phi}^X$). Here, tighter bound estimations offer a considerably larger key rate for a fixed block size, \textcolor{black}{together with optimized basis bias ($P_x$) and pre-attenuation}. \textcolor{black}{Regarding to this calculation, a block size of sent $N_{S,BB84}~=~4\cdot 10^{9}$ signal/pulse, corresponding to $t_S=10^2$ seconds of acqusition time, is used to reach the tolerable loss of 23 dB.  Meanwhile, $t_S=1$ second of acquisition still yields positive key rates at around 17 dB of channel loss, benchmarking the performance of our room-temperature emitter in its simplest form.}

\begin{figure}[!tp]
    \centering
\includegraphics[width=0.49\textwidth]{Fig5_v2.pdf}
    \caption{\textcolor{black}{Attainable SKR in a single-node quantum repeater configuration. Red curves in Fig.~\ref{fig:4} are shown here as horizontal lines. Blue dashed line indicates the point beyond which the repeater configuration starts having advantage over the performance of Ref.~\cite{Zhang.Pan.2024}, whereas the green dashed line shows the limit beyond which the repeater configuration cannot produce a positive key rate.}}
    \label{fig:5}
\end{figure}

\textcolor{black}{\section{Discussion}}

Channel loss is primarily limited by suboptimal single-photon purity, limited collection efficiency of the source, and losses in the experimental setup, all of which can be improved using reported values for defects in hBN. The literature demonstrates that a tunable microcavity significantly narrows the emission spectrum of a defect in hBN and results in a single-photon purity of $g^{(2)}(0) = 0.018$, along with a 10-fold enhancement in photoluminescence~\cite{Vogl.Lam.20190bi}. Additionally, coupling the ZPL emission directly to Alice, without using a PMF (as in our existing setup), will enhance the transmission efficiency by a factor of 2, resulting in a mean photon number before the quantum channel of $\mu = 0.264$. Taking all these improvements into account, our hBN-based room-temperature single-photon source has the potential to reach a SKR of Mbps, approaching the performance of well-established QD sources under similar modulation conditions~\cite{Zhang.Pan.2024}, which, however, require a cryogenic environment, as shown by the blue and black lines in \autoref{fig:4}, respectively.

\autoref{tab:qkd_experiments} lists several reports on PM QKD demonstrations using single-photon sources and their corresponding parameters. The sources are categorized based on their operating conditions: room-temperature (defects in hBN, diamond, GaN, and single molecules) and cryogenic operation (QDs and TMDCs). On the other hand, QKD implementations are categorized based on whether they include active/passive encoding. As observed, room-temperature sources offer a practical implementation with modest SKR, while QD-based sources have the potential for better performance and telecom operation over longer distances through optical fiber, despite their cryogenic challenges.

\begin{table}[!t]  
\centering
\caption{Summary of QKD experiments with various SPS sources}
\resizebox{\columnwidth}{!}{%
\begin{tabular}{%
    >{\raggedright\arraybackslash}p{2.2cm}  
    >{\centering\arraybackslash}p{1.3cm}    
    >{\centering\arraybackslash}p{1.0cm}    
    >{\centering\arraybackslash}p{1.0cm}    
    >{\centering\arraybackslash}p{1.3cm}    
    >{\raggedleft\arraybackslash}p{1.8cm}   
}
    \hline
    \hline
    \multicolumn{1}{l}{\textbf{Single-Photon}} 
      & \multicolumn{1}{c}{\textbf{Clock Rate}} 
      & \multicolumn{1}{c}{\textbf{Active}} 
      & \multicolumn{1}{c}{\textbf{Room}} 
      & \multicolumn{1}{c}{\textbf{QBER}} 
      & \multicolumn{1}{c}{\textbf{Bit-Rate}} \\
    \multicolumn{1}{l}{\textbf{Source, $\lambda_0$ (nm)}} 
      & \multicolumn{1}{c}{\small{(MHz)}} 
      & \multicolumn{1}{c}{\textbf{Encoding}} 
      & \multicolumn{1}{c}{\textbf{Temperature}} 
      & \multicolumn{1}{c}{\small{(\%)}} 
      & \multicolumn{1}{c}{\small{(kbps)}} \\
    \hline
    \hline
    \\*[-0.25cm]
    This Work$^\dagger$, 626 
      & \footnotesize{40} 
      & \footnotesize{$\checkmark$} 
      & \footnotesize{$\checkmark$} 
      & \footnotesize{6.49} 
      & \footnotesize{7 (SKR)} \\
    \hline
    \\*[-0.25cm]
    GaN, 1310 \cite{Zhang.Ling.2024} 
      & \footnotesize{80} 
      & \footnotesize{$\checkmark$} 
      & \footnotesize{$\checkmark$} 
      & \footnotesize{5} 
      & \footnotesize{0.247 (SKR)$^\ast$} \\
    hBN, 671$^\dagger$ \cite{Samaner.Ateş.2022} 
      & \footnotesize{1} 
      & \footnotesize{$\checkmark$} 
      & \footnotesize{$\checkmark$} 
      & \footnotesize{8.95} 
      & \footnotesize{0.24 (SiKR)} \\
    hBN, 650 \cite{Al‐Juboori.Aharonovich.2023} 
      & \footnotesize{0.5} 
      & \footnotesize{$\checkmark$} 
      & \footnotesize{$\checkmark$} 
      & \footnotesize{6} 
      & \footnotesize{0.026 (SKR)} \\
    Mol., 785 \cite{Murtaza.Toninelli.2023} 
      & \footnotesize{80} 
      & \footnotesize{NO} 
      & \footnotesize{$\checkmark$} 
      & \footnotesize{3.4} 
      & \footnotesize{500 (AKR)} \\
    NV, 637 \cite{Leifgen.Benson.2014} 
      & \footnotesize{1} 
      & \footnotesize{$\checkmark$} 
      & \footnotesize{$\checkmark$} 
      & \footnotesize{3} 
      & \footnotesize{2.6 (SKR)} \\
    SiV, 739 \cite{Leifgen.Benson.2014} 
      & \footnotesize{1} 
      & \footnotesize{$\checkmark$} 
      & \footnotesize{$\checkmark$} 
      & \footnotesize{3.2} 
      & \footnotesize{1 (SKR)} \\
    QD, 880 \cite{Waks.Yamamoto.2002vsu} 
      & \footnotesize{76} 
      & \footnotesize{$\checkmark$} 
      & \footnotesize{NO} 
      & \footnotesize{2.5} 
      & \footnotesize{25 (SKR)} \\
    QD, 898 \cite{Heindel.Forchel.2012} 
      & \footnotesize{200} 
      & \footnotesize{$\checkmark$} 
      & \footnotesize{NO} 
      & \footnotesize{3.8} 
      & \footnotesize{35 (SiKR)} \\
    QD, 1545 \cite{Zahidy.Midolo.2024} 
      & \footnotesize{72.6} 
      & \footnotesize{$\checkmark$} 
      & \footnotesize{NO} 
      & \footnotesize{3.25} 
      & \footnotesize{13.2 (SKR)$^{**}$} \\
    QD, 884.5 \cite{Zhang.Pan.2024} 
      & \footnotesize{76.13} 
      & \footnotesize{$\checkmark$} 
      & \footnotesize{NO} 
      & \footnotesize{2.54} 
      & \footnotesize{82 (SKR)***} \\
    QD, 1550 \cite{Morrison.Fedrizzi.2023} 
      & \footnotesize{160.7} 
      & \footnotesize{NO} 
      & \footnotesize{NO} 
      & \footnotesize{2} 
      & \footnotesize{689 (AKR)} \\
    QD, 1556 \cite{Yang.Ding.2024} 
      & \footnotesize{228} 
      & \footnotesize{NO} 
      & \footnotesize{NO} 
      & \footnotesize{0.099} 
      & \footnotesize{68 (AKR)$^{****}$}\\
    TMDC, 807 \cite{Gao.Heindel.2023} 
      & \footnotesize{5} 
      & \footnotesize{NO} 
      & \footnotesize{NO} 
      & \footnotesize{0.69} 
      & \footnotesize{NaN} \\*[0.1cm]
    \hline
\end{tabular}%
}
\\*[0.1cm]
\raggedright \footnotesize{Note: SiKR stands for sifted key rate, AKR for asymptotic key rate, and SKR for secure key rate. [$\dagger$]: based on B92-protocol. [$^\ast$]: for 4.0 dB loss. [$^{**}$]: for 9.6 dB loss. [$^{***}$]: for 15.2 dB loss.  [$^{****}$]: for fiber spool distance of 80 km.}
\label{tab:qkd_experiments}
\end{table}

In order to go beyond point-to-point QKD applications and towards entanglement-based quantum networking, electronic or nuclear spins of such single emitters can be utilized to act as quantum memories and registers~\cite{Bradley.Taminiau.2019, Stas.Lukin.2022, Appel.Atature.2025}. Recent experimental work has shown that emitters in hBN \textcolor{black}{and other 2D materials} possess electronic spins~\cite{Durand.Jacques.2023, Gottscholl.Dyakonov.2021, Stern2022.Atature2024, Stern.Atature.2024}. Furthermore, first-principles calculations predict that some of these defects could exhibit $T_2$ times of around $\sim30$~ms~~\cite{Ye2019, Sajid2022}. Therefore, these spins may become the foundation of quantum registers. In order to understand the required performance for future quantum networking applications, we use a single quantum repeater node~\cite{Langenfeld.Rempe.2021} as a benchmark. This scheme~\cite{Luong.Lutkenhaus.2016} requires at least two identical quantum registers\footnote{Although we assumed a single pair of memories located at the central node, this scheme has been generalized to $m$ pairs that yields a superior performance~\cite{Trenyi2020}.} for entangled state distribution and synchronization. Such a node reduces the effective distance between the communicating parties by half, thereby providing superior performance beyond certain distances. This is illustrated with the red lines in Fig.~\ref{fig:4}, where we plotted the SKR in the asymptotic limit for two different memory times, 5~ms and 10~ms. The reduced slope of the curves up to around 25~dB loss with respect to point-to-point QKD schemes is indicative of the quantum repeater behaviour, which provides better performance over a certain crossover point and better overall loss tolerance. Although we assume the repeater node is placed halfway between Alice and Bob in the low-loss regime, once dephasing becomes significant in the high-loss regime, the optimized position of the repeater node is found to be closer to Bob and at some point becomes fixed. This position optimization helps minimize dephasing errors by providing Alice's quantum memory less time to experience dephasing. As a result, beyond $\sim$25~dB, the optimized key rate scaling behavior changes from $e^{-L / (2L_{\text{att}})}$ to $e^{-L / L_{\text{att}}}$ i.e. direct transition between Alice and Bob~\cite{Luong.Lutkenhaus.2016}. Fig.~\ref{fig:5} shows the SKR as a function of channel loss and memory time. This analysis quantitatively shows that a memory time in the order of $\mathcal{O}(10^{-3})\, \text{s}$ would be sufficient for practical applications. The details of the model and the parameters used for these calculations are given in~\autoref{met:rep}.

In summary, we have demonstrated QKD based on the B92 protocol using single photons from defects in hBN operating at room temperature. Polarization encoding is performed dynamically at a 40 MHz repetition rate using a resonant electro-optic amplitude modulator, resulting in a SiKR of 17.5 kbps and an SKR of 7 kbps with a QBER of 6.49\%. Considering coupling in photonic/plasmonic cavities~\cite{Haußler.Kubanek.2021,Sakib.Shcherbakov.2024,Dowran.Laraoui.2024}, the performance of our source can exceed the fundamental limit of QKD with WCP. For practical quantum communication scenarios over longer distances and satellite-based applications~\cite{Vogl.Lam.20190bi,Abasifard.Vogl.2024}, our results highlight the potential of isolated defects in hBN as an efficient single-photon source compared to other material systems, which can be further miniaturized under electrical excitation~\cite{Yu.Lee.2024, Zhigulin.Aharonovich.2025}.

\vspace{30pt}
\section{Acknowledgements}
This work was supported by the QuantERA II Programme that has received funding from the EU Horizon 2020 research and innovation programme under GA No 101017733 (Comphort), and with funding organisation Scientific and Technological Research Council of Turkey (TUBITAK) under GA nos. 124N115 and 124N110. This work was also supported by the EPSRC Quantum Technology Hub in Quantum Communication (EP/T001011/1), International Network in Space Quantum Technologies (EP/W027011/1), and the Integrated Quantum Networks Research Hub (EP/Z533208/1). M.G. acknowledges funding from the DLR through funds provided by the Federal Ministry for Economic Affairs and Climate Action (Bundesministerium für Wirtschaft und Klimaschutz, BMWK) under Grant No. 50WM2347. The authors thank Serkan Paçal, Çağlar Samaner and Kadir Can Doğan for the fruitful discussions during the preparation of the manuscript and NETES Inc. for their provision of equipment for this research. S.A. acknowledges the support from the Turkish Academy of Sciences (TUBA-GEBIP) and the BAGEP Award of the Science Academy. 

\section{Appendix}
\label{met:met_main}
\subsection{Finite key calculations}
\label{met:met}

QKD parameters used for asymptotic and finite key analysis are presented in \autoref{table:1}. 

\begin{table}[h] 
    \centering 
    \caption{Baseline QKD and security parameters.} 
    \begin{tabular}{>{\centering\arraybackslash}p{5cm}>{\centering\arraybackslash}p{1.5cm}>{\centering\arraybackslash}p{1.5cm}} 
        \hline
        \hline
        Description & Parameter & Value \\
        \hline 
        \hline
        \\*[-0.25cm]
        Excitation (QKD) rate & $R$ & \makecell{\rule{0pt}{1em}40 MHz\\80 MHz$^\triangledown$} \\      \hline
        Collection efficiency & $\mu_{SPS}$ & \makecell{\rule{0pt}{1em}0.052\\0.521$^\triangledown$} \\      \hline
        Transmitter efficiency & $\eta_{tran}$ & \makecell{\rule{0pt}{1em}0.252\\0.507$^\triangledown$} \\     \hline
        Single-photon purity & $g^2(0)$ & \makecell{\rule{0pt}{1em}0.24\\0.018$^\triangledown$} \\     \hline
        Mean-photon number & $\mu_{tran}$ & \makecell{\rule{0pt}{1em}0.0131\\0.264$^\triangledown$} \\     \hline
         Misalignment probability & \textcolor{black}{$P_{mis}$} & \makecell{\rule{0pt}{1em}0.0176\\0.01$^\triangledown$} \\      \hline
        Dark count probability & $p_{dc}$ & \makecell{\rule{0pt}{1em}$8\cdot 10^{-7}$} \\      
        Receiver efficiency & $\eta_{rec}$ & \makecell{\rule{0pt}{1em}$0.42$} \\     
        Privacy amplification failure prob. & $\epsilon_{PA},\epsilon$ & \makecell{\rule{0pt}{1em}$10^{-10}$} \\    
        Smoothing parameter & $\bar{\epsilon}$ & \makecell{\rule{0pt}{1em}$(\epsilon/8)^2$} \\    
        Error reconciliation efficiency & $f_{EC}$ & \makecell{\rule{0pt}{1em}1.16} \\     
        Error reconciliation failure prob. & $\epsilon_{EC}$ & \makecell{\rule{0pt}{1em}$\epsilon$} \\   
        Parameter estimation failure prob. & $\epsilon_{PE}$ & \makecell{\rule{0pt}{1em}$4\epsilon$} \\*[0.1cm]
        \hline
    \end{tabular}
    \\*[0.1cm]
    \raggedright \footnotesize{[$\triangledown$] Parameters used for the improved work~\cite{Vogl.Lam.20190bi}.}
    \label{table:1}
\end{table}

\subsubsection{\textcolor{black}{Finite Key Analysis for B92 protocol}}
\label{met:finite}
\textcolor{black}{To calculate the experimentally achieved secret key rates that are presented in \autoref{fig:3},} we employ the security proof for B92-protocol \cite{Mafu.Petruccione.2013} with smooth-Rényi entropies for the finite-key analysis, such that, length of the secure key fraction after post-processing is bounded by,
\begin{equation}
    \resizebox{0.43\textwidth}{!}{$
    l \leq \displaystyle N_{R}[1-H_{min}(X_A|E)] - L_{EC} - 2 \log_2 \left( \frac{1}{2 \epsilon_{PA}} \right) - \log_2 \left( \frac{2}{\epsilon_{cor}} \right)
    $}
    \label{eq1:main}
\end{equation}
then the protocol is $\epsilon_{qkd}\geq \epsilon_{cor}+\epsilon_{sec}$ secure, if its $\epsilon_{cor}$ correct and $\epsilon_{sec}\geq (2\bar{\epsilon}+\epsilon_{PA})$ secret.

Harnessing the uncertainty relation for smooth-Rényi entropies \cite{Tomamichel.Renner.2010}, Bob's raw key derived from Alice's raw key (A) that conditioned with respect to Eve's uncertainty on the A is quantified as $H_{min}^{\bar{\epsilon}}(X_A|E)$. Further, the information that Bob requires to correct errors utilizing an error reconciliation protocol, is based on Eve's and Bob's uncertainty about A, is also measured in terms of smooth-Rényi entropies $H_{max}^{\bar{\epsilon}}(Z_A|B)$, results the bound of \cite{Tomamichel.Renner.2012};
\begin{equation}
\label{eq:Renyi}
H_{min}^{\bar{\epsilon}}(X_A|E)+ H_{max}^{\bar{\epsilon}}(Z_A|B)\geq q
\end{equation}
where $\bar{\epsilon}\geq 0$ is the smoothing parameter and $q = -\log_2(c)$ is the quality factor quantifies the incompatibility between the measurements $\mathbb{X}^{\otimes n}$ and $\mathbb{Z}^{\otimes n}$ characterized by POVM elements of the non-orthogonal states. Considering perfect qubits prepared and sent by Alice, we set $q=1$. In other words, \autoref{eq:Renyi} is expressed as follows: once Bob achieves the highest accuracy in estimating Alice's raw key in the $Z_A$ basis, Eve's ability to guess Alice's raw key in the $X_A$ basis is minimized, and vice versa. Further, the measure of Bob's uncertainty can only increase if $ H_{max}^{\bar{\epsilon}}(Z_A|B)\leq H_{max}^{\bar{\epsilon}}(Z_A|Z_B)$, for measurement $Z_B$ taken place at Bob which is highly correlated with $Z_A$, such that the maximum uncertainty $H_{max}^{\bar{\epsilon}}(Z_A|Z_B)$ is small and bounded by,
\begin{equation}
\label{eq:Cond1}
H_{max}^{\bar{\epsilon}}(Z_A|Z_B)\leq N_{R}~h(Q)
\end{equation}
where $h(Q)$ is the binary Shannon entropy and -$N_{R}$- is the received noisy key size in bits. Then, minimum entropy conditioned on Eve's knowledge on $X_A$ basis is expressed as, 
\begin{equation}
\label{eq:renyi2}
H_{min}^{\bar{\epsilon}}(X_A|E_{EC})\geq N_{R}~ q-H_{max}^{\bar{\epsilon}}(Z_A|Z_B)
\end{equation}
such that Eve's information on $X_A$ after error reconciliation process is quantified by,

\begin{align}
\label{eq:Cond2}
H_{min}^{\bar{\epsilon}}(X_A|E) &\geq H_{min}^{\bar{\epsilon}}(X_A|E_{EC}) - L_{EC} - \log_2\left(\frac{2}{\epsilon_{cor}}\right) 
\end{align}
where $L_{EC}$ is the amount of information leakage during error reconciliation process. Following the one-way error reconciliation method~\cite{Tomamichel.Elkouss.2017}, the number of leaked bits ($L_{EC}$) is lower bounded by, 
\begin{align}
    L_{EC} &\geq N_R^X \, h(Q) \notag \\
    &+ \left[ N_R^X(1-Q) - F^{-1}(\epsilon_{cor}; N_R^X, 1-Q) \right] \log_2 \left( \frac{1-Q}{Q} \right) \notag \\
    &- \frac{1}{2} \log_2 N_R^X - \log_2 \left( \frac{1}{\epsilon_{cor}} \right) 
    \label{eq:leak}
\end{align}
provided that $F^-1(\epsilon_{cor};N_R^X;1-Q)$ is the inverse of the cumulative distribution of the binomial distribution and $\epsilon_{cor}$ is the correctness failure probability of the reconciliation protocol.

In the context of privacy amplification, Alice and Bob employ a two-universal hash function, utilizing the quantum leftover hash lemma \cite{Tomamichel.Renner.2011}, the final secure key length ($l$) is upper bounded by, 
\begin{equation}
\label{eq:renyi3}
l \leq H_{min}^{\bar{\epsilon}}(X_A|E)-2\log_2(\frac{1}{2\epsilon_{PA}})
\end{equation}
such that combining \autoref{eq:Cond2} and \autoref{eq:renyi3}, final secure key length given in \autoref{eq1:main} is obtained.

\subsubsection{\textcolor{black}{Finite Key Analysis for BB84 protocol}}

\textcolor{black}{We employ the finite-key analysis described by \cite{Morrison.Fedrizzi.2023,Pousa.Jeffers.2024} based on multiplicative Chernoff bounds, to analyse our experimental platform's expected performance in an efficient BB84 scenario, which we plot in \autoref{fig:4}. This method provides} greater key rate for fixed block sizes, by estimating lower bounds of the received non-multiphoton events and upper bound of the phase error rate. Furthermore, lower bounds of received number of signals in key generation ($\mathbb{X}$) and parameter estimation basis ($\mathbb{Z}$) corresponding to non-multiphoton events are defined as, $\displaystyle \underline{N}_{R,nmp}^X = {N}_{R}^X-\bar{N}_{R,mp}^X$ and $\displaystyle \underline{N}_{R,nmp}^Z = {N}_{R}^Z-\bar{N}_{R,mp}^Z$, as the number of received signals are described by $N_R^X = N_S p_X^2 P_{clk}$ and $N_R^Z = N_S p_Z^2 P_{clk}$ respectively, for number of sent signals ($N_S = CR\cdot t_S$), determined by the clock rate and the acquisition time, \textcolor{black}{which is defined as the block size used to distill the key}. Here, employing the Chernoff bound, the upper bound for the received multi-photon events $\bar{N}_{R,mp}^X$ and $\bar{N}_{R,mp}^Z$ are estimated as, $\bar{N}_{R,mp}^{X,Z} =\bar{N}_{R,mp}^{X,Z*}+\Delta^U$, with $\Delta^U=(\beta+\sqrt{8\beta\bar{N}_{R,mp}^{X,Z*}+\beta^2)}/2\bar{N}_{R,mp}^{X,Z*}$, for $\beta=-ln\epsilon_{PE}$, which is bounded by the parameter estimation failure probability ($\epsilon_{PE}$).

For each bases, number of errors ($m_X, m_Z$) are determined by error probability $P_{err}$, expressed in terms of misalignment probability $P_{mis}$ as,
\begin{align}
    P_{err} = \frac{P_0P_{dc}}{2} + P_{dc} + (1-P_{dc})T \mu_{tran} P_{mis} 
\end{align}
given that, $m_X=N_Sp_X^2P_{err}$ and $m_Z=N_Sp_Z^2P_{err}$ for each bases. Considering only parameter estimation basis ($\mathbb{Z}$) is revealed during error correction process, then the phase error rate with received non-multiphoton fraction is estimated as, $\phi^X=m_Z/\underline{N}_{R,nmp}^Z$. On the other hand, for the key generation basis (${N}_{R}^X$), which is never revealed, the phase error rate is upper-bounded by $\bar \phi^X = \phi^X + \gamma^U(N_R^Z,N_R^X,\phi^X,\epsilon) $, such that the function $\gamma^U$ is defined as,  
\begin{equation}
    \scalebox{0.85}{$
    \begin{aligned}
    \gamma^U(n,k,\lambda,\epsilon) &= \frac{1}{2 + 2\frac{A^2G}{(n+k)^2}} \left\{ 
    \frac{(1-2\lambda)AG}{n+k} + \sqrt{\frac{A^2G^2}{(n+k)^2} + 4\lambda(1-\lambda)G} \right\} \\
    A &= \max\{n,k\}, \quad G = \frac{n+k}{nk} \ln \left( \frac{n+k}{2\pi nk\lambda(1-\lambda)\epsilon^2} \right)
    \end{aligned}
    $}
    \label{eq:gammaU}
\end{equation}
where, $ \epsilon = \epsilon_{PA}$. Then the protocol $\epsilon_{QKD}\geq\epsilon_{sec}+\epsilon_{cor}$ is secure, if $\epsilon_{sec}\geq \epsilon_{PA}+\epsilon_{PE}+\epsilon_{EC}$ secret ($ 10^{-10}$) and $\epsilon_{cor}$ correct ($10^{-15}$). Finally, \autoref{eq1:main} can be expressed as, 
\begin{equation}
    \small
    l \leq \displaystyle \underline{N}_{R,nmp}^X[1-h(\bar \phi^X)]-L_{EC}-2log_2(\frac{1}{2\epsilon_{PA}})-log_2(\frac{2}{\epsilon_{cor}})
    \label{eq:cher}
\end{equation}
to estimate the final key fraction, for the key rate $r = l/N_S$.

\subsubsection{Asymptotic Framework}

\textcolor{black}{The asymptotic key rate for the BB84 protocol is given by the Devetak-Winter bound~\cite{devetak2005distillation},
\begin{equation}
\label{eq:Devetak}
    S_\infty \geq S (A|E) - S(A|B),
\end{equation}
where the conditional von Neumann entropies $S(\cdot | \cdot)$ quantify the uncertainty of Alice's subsystem $A$ given the knowledge of Eve's ($E$) and Bob's ($B$) subsystems.}

\textcolor{black}{In the asymptotic limit, an infinite number of transmissions is assumed. Consequently, Bob's count rates converge to their underlying true expectation values. As a result, no Chernoff bound is applied to the expected multi-photon events, hence $\bar{N}_{\mathrm{R,nmp}}^X = \bar{N}_{\mathrm{R,nmp}}^{X,Z*}$. The arbitrarily large block of statistics justifies the following additional simplifications for efficient BB84. First, there is no need to perform parameter estimation ($p_Z \rightarrow 0$), hence all detection events can be used to generate the key ($p_X \rightarrow 1$). Second, in classical postprocessing, we assume perfect error correction efficiency ($f_{EC} \rightarrow 1$), since classical coding theory shows that, with arbitrarily long code blocks, one can approach the Shannon limit arbitrarily closely. Expressing the entropies of Eq. \ref{eq:Devetak} in terms of probabilities and incorporating these asymptotic considerations yields
\begin{equation}
\begin{split}
S_\infty & \geq \lim_{\substack{p_X, \, f_{EC} \, \rightarrow \, 1 \\ p_Z \, \rightarrow \, 0}} {p_{X}^2 P_{clk}}[\Delta (1-h(Q/\Delta))-f_{EC}h(Q)] \\
& \geq {P_{clk}}[\Delta (1-h(Q/\Delta)) - h(Q)].
\end{split}
\label{eq:asym}
\end{equation}
The parameter $\Delta = (P_{clk}-P_m)/P_{clk}$ is calculated using two probabilities. First, the total detection probability $P_{clk} \approx P_{dc} + (1-P_{dc})T\mu_{tran} \eta_{tr}$, where $P_{dc}$ is the dark count probability of the detectors within a single pulse; $T=\eta_{tran}\eta_{Ch}\eta_{rec}$ is the total transmittance (with $\eta_{Ch}$ channel loss and $\eta_{rec}= \eta_{Bob}\cdot\eta_{detector}$ the receiver efficiency); $\mu_{tran}=\mu_{SPS}\cdot \eta_{tran}$ is the mean photon number of the SPS; and $\eta_{tr}$ is the trasmissivity of Alice's attenuator, used to pre-attenuate her SPS before the quantum channel. Second, the multi-photon emission probability, which is upper-bounded by $P_m \leq g^2(0)\mu_{tran}^2 \eta_{tr}^2/2$.} 

Introducing pre-attenuation in Alice's source, defined by the attenuator transmittance $\eta_{tr}$, significantly extends the QKD system's tolerable loss range. In the high-loss regime, dark counts and multi-photon events dominate, the latter depends quadratically on this transmittance, $P_m \alpha \eta_{tr}^2$, whereas the detection probability scales linearly, $P_{clk} \alpha \eta_{tr}$. Consequently, the multi-photon probability decreases more rapidly, enhancing the key generation rate.

For the simulations, the expected QBER ($Q$) is defined by,
\begin{equation}
Q =\frac{P_{mis}T\mu_{tran}+P_{dc}/2}{P_{clk}}
\end{equation}
where the parameter -$P_{mis}$- represents the misalignment probability, described as static error contribution due to component imperfections.

\subsection{\textcolor{black}{Single quantum repeater node calculations}}
\label{met:rep}

\textcolor{black}{The scheme analyzed in Ref.~\cite{Luong.Lutkenhaus.2016} relies on two quantum registers, or memories (QM-A and QM-B),} \textcolor{black}{each capable of creating a photon-memory entangled state. A photon entangled with QM-A is prepared (with efficiency $\eta_p$ and over time $T_p$) and sent to Alice, who performs a BB84 measurement on it, repeating the process until she successfully detects a photon. The same procedure is then carried out with Bob and QM-B. Once both photons have been measured, a Bell measurement is performed on the two QMs, and the result is shared with Bob. Depending on the outcome, Bob may need to apply a bit flip to his BB84 measurement result} \textcolor{black}{to obtain the same bit as Alice,} \textcolor{black}{—specifically, if he measured in the \(Z\) basis and the Bell measurement resulted in \(|\psi^+\rangle\) or \(|\psi^-\rangle\), or if he measured in the \(X\) basis and the Bell measurement resulted in \(|\phi^-\rangle\) or \(|\psi^-\rangle\). Table~\ref{table:3} shows the  relevant parameters used in the calculations that are shown in Figs.~\ref{fig:4} and \ref{fig:5}. The position of this central node with respect to Alice and Bob can be optimized for a given channel loss to maximize the achievable key rate. Exact formulae for these calculations are not reproduced here and can be found in Ref.~\cite{Luong.Lutkenhaus.2016}.}

\begin{table}[h] 
    \centering 
    \caption{Parameters used in MA-QKD calculations} 
    \begin{tabular}{>{\centering\arraybackslash}p{5cm}>{\centering\arraybackslash}p{1.5cm}>{\centering\arraybackslash}p{1.5cm}} 
        \hline
        \hline
        Description & Parameter* & Value \\
        \hline 
        \hline
        \\*[-0.25cm]
        Entangled state preparation efficiency & $\eta_p$ & 0.7 \\
        Preparation time & $T_{p}$ & $10^{-6}$~s\\ [0.1cm]
        Fiber coupling and frequency conversion efficiency & $\eta_c$ & 0.7 \\
        Detection efficiency & $\eta_{d}$ & 0.7 \\
        BSM ideality parameter& $\lambda_{BSM}$ & 1 \\
        BSM success probability  & $\eta_{BSM}$ & 0.175 \\
        Error correction inefficiency & $f$ & 1.16\\
       Misalignment errors & $e_{mA}$, $e_{mB}$ & $10^{-2}$ \\ 
        Memory time & $T_2$ & varied \\ 
    Attenuation length & $L_{att}$ & 22~km \\

        \hline
    \end{tabular}
    \\*[0.1cm]
    \raggedright \footnotesize{*We adopt the same parameter notation as in Ref.~\cite{Luong.Lutkenhaus.2016} for consistency.} 

    \label{table:3}
\end{table}

\bibliographystyle{apsrev4-1}
\bibliography{main}
\end{document}